\documentclass[aps,prb,twocolumn,showpacs,superscriptaddress,amsmath,amssymb]{revtex4-2}

\usepackage{float} 
\usepackage{amsmath}
\usepackage{amsfonts}
\usepackage{amssymb}
\usepackage{amsthm}
\usepackage{graphicx}
\usepackage[colorlinks=true,linkcolor=blue,citecolor=blue,urlcolor=blue]{hyperref}
\usepackage{color}
\usepackage{flafter}
\usepackage[normalem]{ulem}

\usepackage[toc,page]{appendix}
\usepackage{makecell}
\usepackage{slashed}
\usepackage{tikz-cd}
\usepackage{makecell}
\usepackage{graphicx}

\allowdisplaybreaks

\definecolor{OliveGreen}{rgb}{0,0.6,0}
\definecolor{Orange}{rgb}{1.00, 0.65, 0}
\definecolor{Grey}{rgb}{0.43, 0.5, 0.5}

\newcommand{\Eq}[1]{Eq.(\ref{#1})}
\newcommand{\nn}{\nonumber\\}

\newcommand{\ket}[1]{|{#1}\rangle}

\newcommand{\be}{\begin{eqnarray}}
\newcommand{\ee}{\end{eqnarray}}
\newcommand{\bpm}{\begin{pmatrix}}
\newcommand{\epm}{\end{pmatrix}}

\newcommand{\comment}[1]{}

\newcommand{\braket}[3]{\langle#1|#2|#3\rangle}

\begin{document}

\title{Pair-breaking scattering interference as a mechanism for superconducting gap modulation}
\author{Zhi-Qiang Gao}
\affiliation{Department of Physics, University of California, Berkeley, California 94720, USA}
\author{Yu-Ping Lin}
\affiliation{Department of Physics, University of California, Berkeley, California 94720, USA}
\author{Dung-Hai Lee}
\email{dunghai@berkeley.edu}
\affiliation{Department of Physics, University of California, Berkeley, California 94720, USA}
\affiliation{Materials Sciences Division, Lawrence Berkeley National Laboratory, Berkeley, California 94720, USA}

\begin{abstract}
We propose the ``pair-breaking scattering interference" as a general source of coherence peak modulations in superconductors. Assuming this mechanism, we present a simple physical picture for the coherence peak modulations in overdoped cuprate Bi$_2$Sr$_2$Ca$_2$Cu$_3$O$_{10+\delta}$ (Bi-2223), ferromagnetic iron pnictide EuRbFe$_4$As$_4$ (Eu-1144), and kagome metals $A$V$_3$Sb$_5$ ($A=$ K, Rb, and Cs). Specifically, we explain the wavevectors, the particle-hole symmetry, and the dependence on the internal or external Zeeman-field of the coherence peak modulations. This work is intended as a cautious reminder to the scientific community when asserting the existence of a pair density wave phenomenon in the absence of tunneling conductance modulations in the normal state.

\end{abstract}

\maketitle

\section{ Introduction} The pair density wave (PDW), a superconducting (SC) state characterized by the condensation of Cooper pairs possessing a non-zero center-of-mass momentum \cite{FF,LO}, has been a sought-after realm of inquiry.  Owing to the advance of scanning tunneling microscope (STM) techniques, experimental observations relevant to PDW can be made by detecting the spatial modulations of either the height or the energy positions (the SC gap) of the coherence peaks. For example, uni-directional modulations in the height and the energy positions of the coherence peaks have been observed in the halo of vortex cores for Bi$_2$Sr$_2$CaCu$_2$O$_8$ (Bi-2212) \cite{Edkins}. Bi-directional coherence peak modulations have been recently observed in overdoped Bi$_2$Sr$_2$CaCu$_2$O$_{8+\delta}$ (Bi-2212) and   Bi$_2$Sr$_2$Ca$_2$Cu$_3$O$_{10+\delta}$ (Bi-2223) (where the modulation wavevector $\approx (\pi,\pi)$) \cite{Zou2022}, heavy-fermion superconductor UTe$_2$ \cite{gu23n}, and kagome metal $A$V$_3$Sb$_5$ with $A=$ K, Rb, and Cs \cite{chen21n,Deng2024Na,Deng2024NM}. Adding to this list, uni-directional modulations of the SC gap have been reported for the ferromagnetic (FM) iron pnictide EuRbFe$_4$As$_4$ (Eu-1144) \cite{zhao23n}, and along the defect domain walls formed when the monolayer Fe(Se,Te) is grown on SrTiO$_3$ \cite{liu23n}.

This paper excludes the discussion of coherence peak modulations that could be attributed to a secondary order arising from the mere coexistence of a normal-state charge density wave (CDW) and a uniform SC order. The observed ``PDW" in UTe$_2$ falls within this category. In contrast, the coherence peak modulations in Bi-2212, Bi-2223, Eu-1144, and $A$V$_3$Sb$_5$ with $A=$ K, Rb, and Cs belong to the category we shall discuss, where density wave orders are absent in the normal states of these materials.

It is very important to note that the concept to be expounded below was initially introduced in Ref.~\cite{Zou2022} for Bi-2212 and Bi-2223, and in this paper, we extend its applicability to encompass the cases of Eu-1144 and $A$V$_3$Sb$_5$. We have coined this concept the ``pair-breaking scattering interference" (PBSI), to differentiate it from the PDW phenomenon. This work serves as a cautious reminder to the scientific community when asserting the existence of a PDW under the situation that a similar modulation in the local density of states is absent in the normal state. We acknowledge that authentic PDW in many materials other than Bi-2212, Bi-2223, Eu-1144, and kagome metals can indeed exist \cite{Spalek2022}. Under such conditions our formalism does not apply. Thus, we reiterate that the goal of this work is to propose and illustrate an alternative mechanism for coherence peak modulations other than PDW, and not all coherence peak modulations can fit into our formalism.

The PBSI formalism relies on the existence of a \textit{uniform spin-singlet} SC order. However, the microscopic origin of such uniform SC order is not in the scope of this work. Instead, we adopt a BdG-type Hamiltonian to describe it. Even in strongly correlated superconductors such as cuprates, BdG Hamiltonian remains an accurate description for the superconducting quasiparticle dynamics, which is verified by numerous experiments. Within this framework, we show that coherence peak modulations can emerge as a consequence of the PBSI. This serves as an alternative mechanism alongside PDW explaining the phenomena of coherence peak modulations.

\begin{figure}[t]
  \centering
		\includegraphics[width=0.95\linewidth]{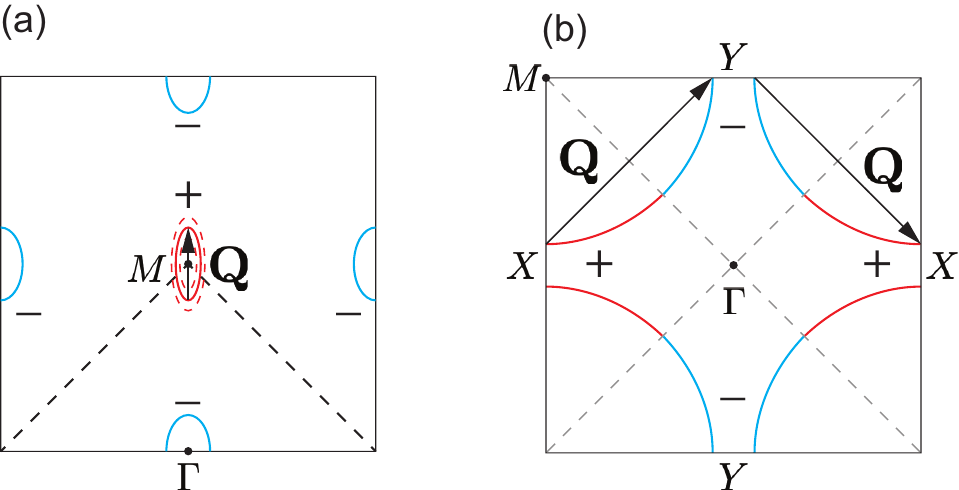}
		\caption{Fermi surfaces, gap function signs, and impurity-scattering wavevectors of (a) Eu-1144 and (b) Bi-2223. For clarity, the BZ of Eu-1144 is drawn with the center at $M$, and the BZ edge is indicated by the black dashed lines. The Fermi surfaces are colored, where the red and blue segments have positive and negative gap functions (labeled with $\pm$), respectively. The red dashed curves in panel (a) are equal-energy contours of the SC states. The gray dashed lines in (b) indicate the gap-vanishing axes of $d$-wave pairing. The arrows $\mathbf{Q}$ represent the wavevectors of the coherence peak modulations.}\label{fig:fs}
\end{figure}

\section{ The experimental phenomenology} 
To begin, we briefly summarize the phenomenon of the coherence peak modulation arsing from the spatial modulation of the superconducting gap. As a function of spatial coordinate $\mathbf{r}$, the positive and negative coherence peaks detected at the SC gap edges are situated at energies
\begin{equation}
\begin{aligned}
\Delta_+(\mathbf{r})&=\Delta_0+\Delta_1\cos(\mathbf{Q}\cdot\mathbf{r}),\\
\Delta_-(\mathbf{r})&=-[\Delta_0+\rho\Delta_1\cos(\mathbf{Q}\cdot\mathbf{r})].
\end{aligned}
\end{equation}
Here $2\Delta_0>0$ denotes the average energy distance between the coherence peaks, and $\Delta_1\cos(\mathbf{Q}\cdot\mathbf{r})$ represents the modulation component with amplitude $\Delta_1>0$ (usually small compared with $\Delta_0$) at wavevector $\mathbf{Q}$. The coefficient $\rho=\pm 1$ controls the particle-hole symmetry (asymmetry) nature of the modulation. In Eu-1144, the modulation is uni-directional and particle-hole symmetric, hence $\rho=1$ \cite{zhao23n}. Meanwhile, Bi-2212 and Bi-2223 exhibits bi-directional, ${\bf Q}\approx (\pi,\pi)$, particle-hole asymmetric modulations with $\rho=-1$ \cite{Zou2022}. To understand how the coherence peak modulations arise, we look into the electronic structure and the pairing symmetry associated with the uniform SC order in these materials (Fig.~\ref{fig:fs}).

\subsection{Eu-1144} The Fermi surfaces are composed of hole pockets and electron pockets at the Brillouin zone (BZ) center $\Gamma$ and corner $M$, respectively \cite{hemmida21prb}. The SC state is nodeless, and the widely believed pairing symmetry is $s_\pm$ (meaning the signs of the gap function are opposite between the $\Gamma$- and $M$-point Fermi pockets). The observed SC gap modulation (with wavevector $\sim$ 8 lattice constant) roughly agrees with $2\pi$ over the momentum distance between the tips of the ellipsoidal electron pockets centered at M [see Fig.~\ref{fig:fs} (a)]. We attribute the lack of the $C_4$ rotation symmetry to the nematic order pervasive in iron pnictides.
In the subsequent discussions we \textit{assume} the SC gap modulation is due to the PBSI of the Bogoliubov quasi-particles (QPs) at the tips of the electron pockets. This assumption is reasonable because the area enclosed by SC state equal-energy contours [red dashed curves in Fig.~\ref{fig:fs}(a)] expands the fastest at the tips of ellipsoid, and the density of states is highest there. We refer to the tips of the ellipsoid as ``hot spots''. 

\subsection{Bi-2212 and Bi-2223}  The Fermi surfaces of the overdoped Bi-2212 and Bi-2223 are schematically shown in Fig. \ref{fig:fs} (b). The different colors indicate that the gap function has opposite signs on the respective segments of the Fermi surface. The superconducting gap is maximal on the Fermi surfaces closest to X and Y. Moreover, due to the proximity of the van Hove singularity, the density of states is the largest, and hence the hot spots are situated there.  The bi-directional coherence peak-modulation wavevectors $\mathbf{Q}\approx(\pi,\pi)$ \cite{Zou2022} again approximately fit the wavevectors connecting the inequivalent hot spots.

\section{The PBSI mechanism for the periodic SC gap modulation} 
Consider the BdG-type Hamiltonian describing a uniform spin-singlet superconductor
\begin{equation}
H=\sum_\mathbf{k}\Psi_\mathbf{k}^\dag(\epsilon_\mathbf{k}ZI-hZZ-\Delta^R_\mathbf{k}YY-\Delta^I_\mathbf{k}XY) \Psi_\mathbf{k}.\label{eq:H}
\end{equation}
Spin-triplet superconductors such as UTe$_2$ are not included in our discussion. The Nambu spinor at momentum ${\bf k}$ is defined as $\Psi_\mathbf{k}=(c_{\mathbf{k}\uparrow},c_{\mathbf{k}\downarrow},c_{-\mathbf{k}\uparrow}^\dag,c_{-\mathbf{k}\downarrow}^\dag)^T$. $I$, $X$, $Y$, $Z$ are the Pauli matrices. Two Pauli matrices standing next to each other, e.g., XY, represents the tensor product, with the first Pauli matrix acting in the particle-hole space and the second one acting in the spin space.

The dispersion relation $\epsilon_\mathbf{k}$ is defined with respect to the Fermi level $\epsilon_{\mathbf{k}_F}=E_F=0$, where the Fermi momenta $\mathbf{k}_F$ constitute the Fermi surface. Meanwhile, the gap function $\Delta_\mathbf{k}=\Delta^R_\mathbf{k}+i\Delta^I_\mathbf{k}=|\Delta_\mathbf{k}|e^{i \theta_\mathbf{k}}
$ is included to capture the Cooper pairing. A Zeeman term with $h>0$ is also introduced in \Eq{eq:H}, which accounts for the coupling to a spontaneous generated internal field due to ferromagnetism or an external Zeeman field. Since the Zeeman term $\sim ZZ$ commutes with the Hamiltonian $H$, the spins $\uparrow$ and $\downarrow$ are the good quantum numbers with $ZZ=+1$ and $-1$, respectively. Thus, the Hamiltonian in \Eq{eq:H} decomposes into two spin parts
\be
&&H_{\uparrow}=\Psi_{\mathbf{k}\uparrow}^\dag(\epsilon_\mathbf{k}Z-hI+\Delta^R_\mathbf{k}X-\Delta^I_\mathbf{k}Y)\Psi_{\mathbf{k}\uparrow},\nn&&H_{\downarrow}=\Psi_{\mathbf{k}\downarrow}^\dag(\epsilon_\mathbf{k}Z+hI-\Delta^R_\mathbf{k}X+\Delta^I_\mathbf{k}Y)\Psi_{\mathbf{k}\downarrow},
\label{eq:Hpm}
\ee
with $\Psi_{\mathbf{k}\uparrow}=(c_{\mathbf{k}\uparrow},c_{-\mathbf{k}\downarrow}^\dag)^T$ and $\Psi_{\mathbf{k}\downarrow}=(c_{\mathbf{k}\downarrow},c_{-\mathbf{k}\uparrow}^\dag)^T$.

\begin{figure}[t]
  \centering
		\includegraphics[scale=0.6]{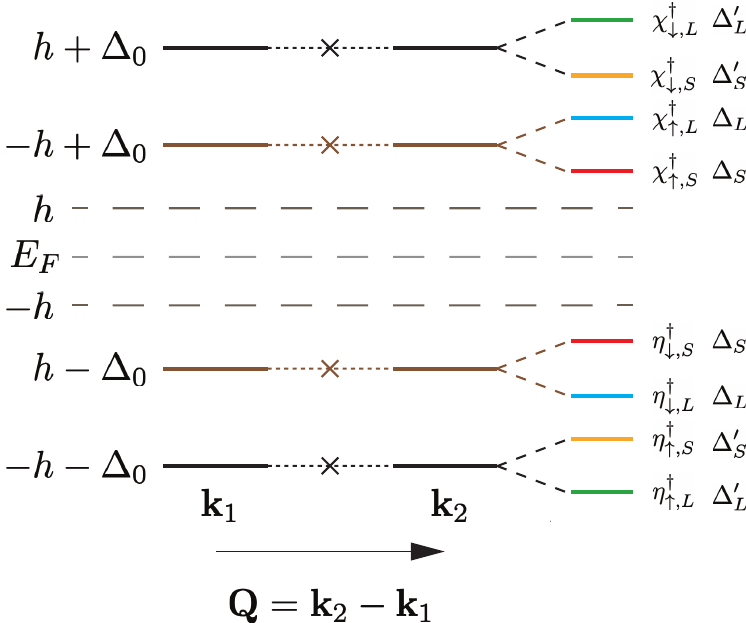}
		\caption{Evolution of quasiparticle energy levels under Zeeman field and the pair-breaking scattering interference (PBSI). In the ideal superconducting state the quasiparticles exhibit particle-hole symmetric energy levels $\pm\Delta_0$ at the two hot spots $\mathbf{k}_{1,2}$. The Zeeman field $h$ breaks the spin degeneracy. The two brown energy levels $\pm(-h+\Delta_0)$ are most important to the physics at the gap edge. In the presence of PBSI (the crossed dashed lines), the energy levels at the two hot spots get hybridized and split. The original two-fold degenerate black energy levels are split into green and orange levels, while brown levels are split into red and blue ones.}\label{fig:energy}
\end{figure}

The eigenvalues of the BdG Hamiltonians $H_{\uparrow,\downarrow}$ are symmetrically centered about energies $\epsilon=\mp h$. At the Fermi momenta $\mathbf{k}_F$, the QP operators are 
\begin{equation}
\begin{aligned}
\chi_{\mathbf{k}_F\uparrow,\downarrow}&=\frac{1}{\sqrt{2}}(c_{\mathbf{k}_F\uparrow,\downarrow}\pm e^{i\theta_{\mathbf{k}_F}}c_{-\mathbf{k}_F\downarrow,\uparrow}^\dag),\\
\eta_{\mathbf{k}_F\uparrow,\downarrow}&=\frac{1}{\sqrt{2}}(c_{\mathbf{k}_F\uparrow,\downarrow}\mp e^{i\theta_{\mathbf{k}_F}}c_{-\mathbf{k}_F\downarrow,\uparrow}^\dag).
\end{aligned}
\label{eq:bogo}
\end{equation}
Assuming a small Zeeman splitting $h<|\Delta_{\mathbf{k}_F}|$ \cite{note1}, the QP energy levels associated with  $\chi_{\mathbf{k}_F\uparrow,\downarrow}$, with the energies $|\Delta_{\mathbf{k}_F}|\mp h>0$, always sit above the Fermi level (Fig.~\ref{fig:energy}). Conversely, the QP energy levels associated with $\eta_{\mathbf{k}_F\uparrow, \downarrow}$ carrying the energies $-|\Delta_{\mathbf{k}_F}|\mp h<0$ stay below the Fermi level. Note that the two QP levels associated with $\chi_{\mathbf{k}_F\uparrow}$ and $\eta_{\mathbf{k}_F\downarrow}$ have energies closest to the Fermi level, namely, $\pm (|\Delta_{\mathbf{k}_F}|-h)$. They are the ones most relevant to the gap-edge physics.

Now we study the effect of impurity scattering on the Bolgoiubov QP levels associated with the two hot spots $\mathbf{k}_1$ and $\mathbf{k}_2$. In the following we define $\mathbf{Q}=\mathbf{k}_1-\mathbf{k}_2$. The gap functions are $\Delta_{\mathbf{k}_1}=s\Delta_{\mathbf{k}_2}=\Delta_0e^{i\theta}$, where $\Delta_0>0$ and $s=\pm1$ represents the relative sign of the gap function at the hot spots. At this juncture, it is important to note that we {\it do not} assume  the impurity scattering can only impart momentum $\mathbf{Q}$ to the quasiparticles. The fact that we ignore all other momentum-transfer scatterings, except those between the tips of the equal energy contours in Eu-1144 [see Fig. \ref{fig:fs} (a)] or anti-nodal regions [see Fig. \ref{fig:fs} (b)], is because of the large joint density of states between the hot spots \cite{Wang2003}. In addition, because the superconducting gap is maximal at the hot spots, they control the coherence peaks. It is also important to note that the PBSI mechanism aims to explain observations of \textit{local} experimental probes, such as STM, under a fixed configuration of impurities with a moderate density. In this case, the multi-scattering effects by different impurities, including disorder average, do not wash out the scattering interference completely. Meanwhile, the finite density of impurities ensures the occurrence of observable modulations. In addition, phenomena such as coherence peak modulation and quasiparticle interference are absent if the disorder average is performed. Hence a disorder-averaged theory is not relevant for our purposes. This approach, under assumptions stated above, has proven effective in capturing relevant physics in superconductors \cite{Wang2003}.

Associated with $\chi_{\mathbf{k}_{1,2}\uparrow\downarrow}$ there are four QP levels above the Fermi energy $E_F$, and associated with $\eta_{\mathbf{k}_{1,2}\uparrow\downarrow}$ there are four QP levels below the Fermi level. In the following we shall ignore the impurity induced mixing between the $\chi$ levels and the $\eta$ levels, while only focus on the mixing among the levels below or above $E_F$. The reason for doing so is that such mixing has the effect of reducing the superconducting gap, hence is pair-breaking. Without affecting the nature of magnetic scattering we shall assume it preserves the $z$-component of the spin, under which the impurity-induced pair-breaking (PB) Hamiltonians for spin-up ($\uparrow$) and spin-down ($\downarrow$) read
\begin{equation}
\begin{aligned}
H_{\text{PB},\uparrow\downarrow}
&=\bigg[V^\text{s}_\mathbf{Q}\bigg(\frac{1-s}{2}\bigg)\pm V^\text{m}_\mathbf{Q}\bigg(\frac{1+s}{2}\bigg)\bigg]\\&\quad\times(\chi_{\mathbf{k}_1\uparrow\downarrow}^\dag
\chi_{\mathbf{k}_2\uparrow\downarrow}+\eta_{\mathbf{k}_1\uparrow\downarrow}^\dag\eta_{\mathbf{k}_2\uparrow\downarrow})+\text{h.c.}.
\end{aligned}
\end{equation}
Here $V^\text{s}_\mathbf{Q}$ and $V^\text{m}_\mathbf{Q}$ denote the scalar and magnetic impurity potentials, respectively. See Supplemental Material (S. M.) \cite{supp} Section I for a detailed derivation of this Hamiltonian. Importantly, (i) if the signs of the gap function at the hot spots are the same ($s=1$), the scalar impurity does not break pairs. This is consistent with Anderson's theorem. In contrast, the magnetic impurity breaks Cooper pairs. On the other hand, if (ii) the gap sign at the hot spots are opposite ($s=-1$), the roles of the scalar and magnetic impurities are switched. Eu-1144, and overdoped Bi-2212 and Bi-2223, precisely correspond to the two cases (i) and (ii) above, respectively.

\begin{figure}[t]
  \centering
		\includegraphics[scale=0.37]{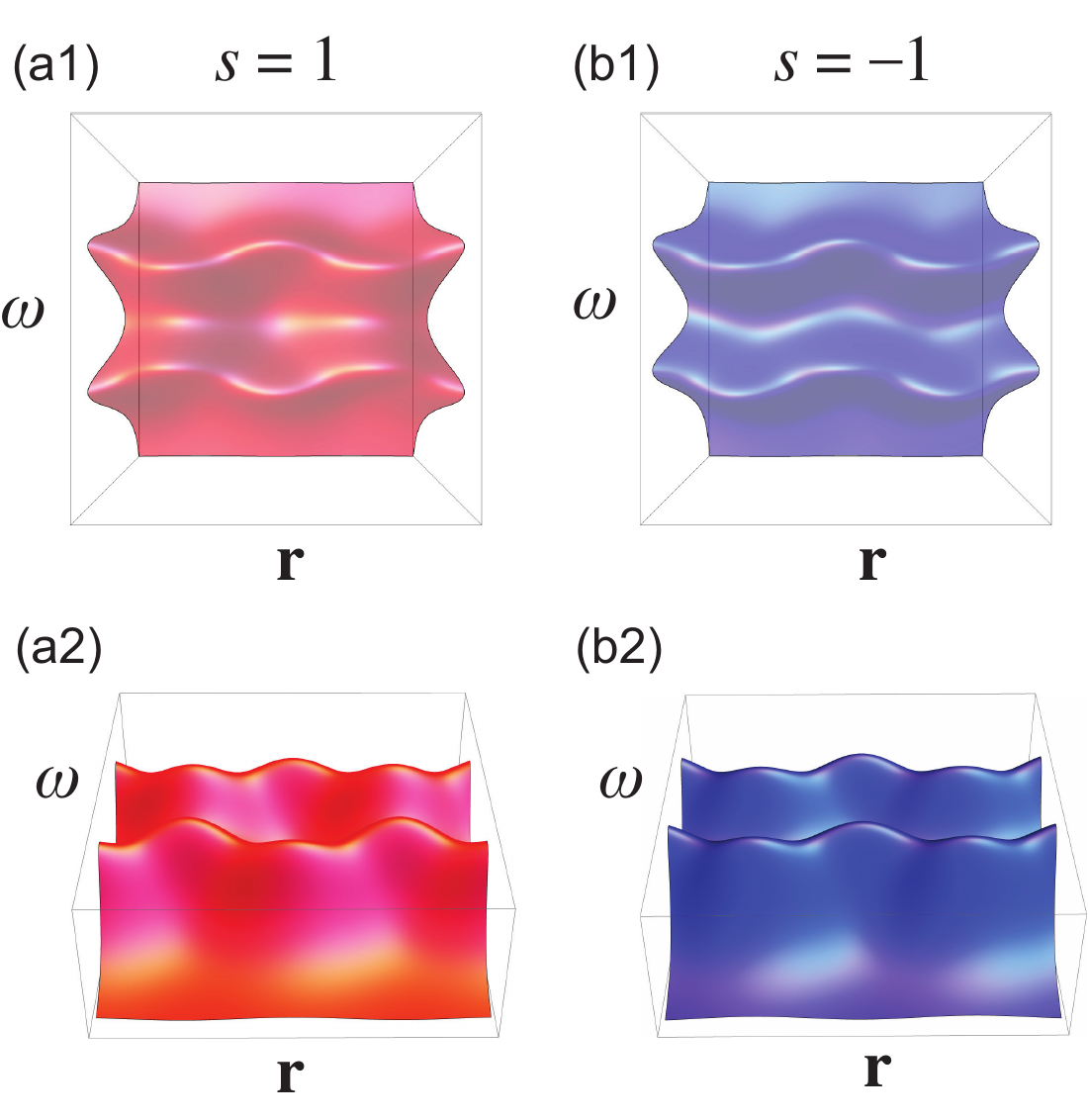}
		\caption{The coherence peak modulations derived from the energy levels at the hot spots, shown by the LDOS (label omitted for clarity) as functions of spatial position $\mathbf{r}$ and energy bias $\omega$. Here we use the Lorentzian function representing the contribution of each energy level to the local density of states. (a1) When the gap function has the same sign at the two hot spots ($s=1$), such as in Eu-1144, the modulations are particle-hole symmetric. (b1) If the gap function has opposite signs at the two hot spots ($s=-1$), as in the case of Bi-2223, the modulations are particle-hole asymmetric. From (a2) and (b2), it can be seen that the height of coherence peaks also modulates for both $s=1$ and $s=-1$ cases, and shows a maximum when the modulation amplitude is reached.} \label{fig:ldos}
\end{figure}

In the presence of the pair-breaking scattering, the QP states at the hot spots are mixed. This mixing induces the splitting of QP energy levels (Fig.~\ref{fig:energy}). For weak scattering such splitting is small compared with the other energy scales such as $h$ and $\Delta_0$. Defining the pair-breaking impurity-scattering potential as $V^\text{s/m}_\mathbf{Q}=Ve^{i\phi}$ with magnitude $V>0$ and phase $\phi$, the new QP operators are
\begin{equation}
\begin{aligned}
\chi_{\uparrow,S}&=\chi_{\mathbf{k}_1\uparrow}-e^{i\phi}\chi_{\mathbf{k}_2\uparrow},\quad
\chi_{\uparrow,L}=\chi_{\mathbf{k}_1\uparrow}+e^{i\phi}\chi_{\mathbf{k}_2\uparrow},
\\
\chi_{\downarrow,S}&=\chi_{\mathbf{k}_1\downarrow}+se^{i\phi}\chi_{\mathbf{k}_2\downarrow},\quad
\chi_{\downarrow,L}=\chi_{\mathbf{k}_1\downarrow}-se^{i\phi}\chi_{\mathbf{k}_2\downarrow},
\\
\eta_{\uparrow,S}&=\eta_{\mathbf{k}_1\uparrow}+e^{i\phi}\eta_{\mathbf{k}_2\uparrow},\quad
\eta_{\uparrow,L} =\eta_{\mathbf{k}_1\uparrow}-e^{i\phi}\eta_{\mathbf{k}_2\uparrow},
\\
\eta_{\downarrow,S}&=\eta_{\mathbf{k}_1\downarrow}-se^{i\phi}\eta_{\mathbf{k}_2\downarrow},\quad
\eta_{\downarrow,L}=\eta_{\mathbf{k}_1\downarrow}+se^{i\phi}\eta_{\mathbf{k}_2\downarrow}.
\end{aligned}
\label{eq:SL}
\end{equation}
Here $S$ and $L$ label the eigenstates with smaller and larger energy gaps from the Fermi level, respectively. The $1/\sqrt{2}$ factors are neglected for clarity. Among these eigenstates, the group of four corresponding to $\chi_{\uparrow,S,L}$ and $\eta_{\downarrow,S,L}$ (see Fig.~\ref{fig:energy}) are the most relevant to physics at the gap edge.

Since the eigenstates associated with the QP operators in \Eq{eq:SL} are linear combinations of those at the two different hot-spot momenta, the absolute value square of them, which appears in the expression of the local density of states (LDOS), will have a cross term that modulate at the momentum ${\bf Q=k_1-k_2}$. This is the PBSI we refer to earlier.

Let us first analyze the limit where we can completely ignore the four QP levels farther from the Fermi energy.
At spatial position $\mathbf{r}$ and energy bias $\omega$, we derive the LDOS from the four low-energy eigenstates corresponding to $\chi_{\uparrow,S,L}$ and $\eta_{\downarrow,S,L}$. See S. M. \cite{supp} Section II for a detailed derivation. The LDOS at positive bias and negative bias are 
\begin{equation}
\begin{aligned}
D_+(\omega,\mathbf{r})&\sim\left[1-\cos(\mathbf{Q}\cdot\mathbf{r}+\phi)\right]\delta(\omega-\Delta_S)\\
&+\left[1+\cos(\mathbf{Q}\cdot\mathbf{r}+\phi)\right]\delta(\omega-\Delta_L),\\
D_-(\omega,\mathbf{r})&\sim\left[1-s\cos(\mathbf{Q}\cdot\mathbf{r}+\phi)\right]\delta(\omega+\Delta_S)\\
&+\left[1+s\cos(\mathbf{Q}\cdot\mathbf{r}+\phi)\right]\delta(\omega+\Delta_L),
\label{eqD}
\end{aligned}
\end{equation}
where $\Delta_{L,S}=\Delta_0-h\pm V$. In \Eq{eqD} the proportionality constant is the tunneling matrix element. For $s=1$ (the case of Eu-1144), the coherence peaks exhibit particle-hole symmetric modulations [Fig.~\ref{fig:ldos}(a)], consistent with the experimental  observation \cite{zhao23n}.  On the other hand, for $s=-1$ (the case of Bi-2223 and Bi-2212) the coherence peaks exhibit particle-hole asymmetric modulation [Fig.~\ref{fig:ldos}(b)] as seen in Ref.~\cite{Zou2022}.

Now we take all QP levels into account. In Eu-1144, the coherence peak modulations only occur below a FM transition temperature which is lower than the SC $T_c$ \cite{zhao23n}.
In contrast, in Bi-2212 and Bi-2223, the coherence peak appears in the absence of the magnetic field or spontaneous magnetization \cite{Zou2022}. To understand the difference in these two cases, we inspect how the QP states corresponding to $\chi_{\downarrow,S,L}$ and $\eta_{\uparrow,S,L}$ enter the coherence peak modulations. The LDOS contributed by these states are
\begin{equation}
\begin{aligned}
D_+'(\omega,\mathbf{r})&\sim[1+s\cos(\mathbf{Q}\cdot\mathbf{r}+\phi)]\delta(\omega-\Delta_S')\\&+[1-s\cos(\mathbf{Q}\cdot\mathbf{r}+\phi)]\delta(\omega-\Delta_L'),\\
D_-'(\omega,\mathbf{r})&\sim[1+\cos(\mathbf{Q}\cdot\mathbf{r}+\phi)]\delta(\omega+\Delta_S')\\&+[1-\cos(\mathbf{Q}\cdot\mathbf{r}+\phi)]\delta(\omega+\Delta_L'),
\end{aligned}
\end{equation}
where $\Delta'_{L,S}=\Delta_0+h\pm V$. Again, the proportionality constant is the tunneling matrix element.
Assuming that $D_\pm$ and $D'_\pm$ contribute with equal weight, the total LDOS is given by the $$D_{{\rm tot},\pm}(\omega,\mathbf{r})=D_{\pm}(\omega,\mathbf{r})+D'_{\pm}(\omega,\mathbf{r}).$$
When the Zeeman field is absent $h=0$,  $\Delta_{S,L}'=\Delta_{S,L}$, yielding the total LDOS given by
\be
&&D_{\rm tot,+}(\omega,\mathbf{r})
\sim\Big\{\left[1+\frac{s-1}{2}\cos(\mathbf{Q}\cdot\mathbf{r}+\phi)\right]\delta(\omega-\Delta_S)\nn&&+\left[1-\frac{s-1}{2}\cos(\mathbf{Q}\cdot\mathbf{r}+\phi)\right]\delta(\omega-\Delta_L)\Big\},\nn&&D_{\rm tot,-}(\omega,\mathbf{r})\sim\Big\{\left[1-\frac{s-1}{2}\cos(\mathbf{Q}\cdot\mathbf{r}+\phi)\right]\delta(\omega+\Delta_S)\nn&&+\left[1+\frac{s-1}{2}\cos(\mathbf{Q}\cdot\mathbf{r}+\phi)\right]\delta(\omega+\Delta_L)\Big\}.
\label{eq:LDOS}
\ee
The modulations vanish in the case $s=1$ (Eu-1144), due to the destructive interference between $D_\pm$ and $D'_\pm$. This explains the absence of coherence peak modulations above the FM transition temperature in Eu-1144 \cite{zhao23n}, where the spontaneously generated Zeeman field is zero. The  spin-splitting Zeeman field, wherever it originates from, serves as a minimal requirement for the modulations to occur in the case $s=1$. A clarification is in order, namely, where the magnetic scattering of QP comes from in Eu-1144. Below the ferromagnetic transition temperature, where a spontaneous magnetization breaking the time reversal symmetry exists, any impurity scattering should acquire some magnetic nature. On the other hand, $D_\pm$ and $D'_\pm$ constructively interfere in the case $s=-1$. Therefore, the coherence peak modulations exist without Zeeman field, which is consistent with the experiments of Bi-2212 and Bi-2223 \cite{Zou2022}.

So far, we have studied two extreme cases: (i) the high-energy QP eigenstates are completely ignored,  and (ii) the high and the low energy QP eigenstates contribute with equal weight. In reality, the total LDOS under Zeeman field $h\neq0$ should be somewhere in-between. In the $s=1$ case (applicable to Eu-1144), the total LDOSs are
\begin{equation}
\begin{aligned}
D_{{\rm tot},\pm}^{s=1}(\omega,\mathbf{r})&\sim[1-W(h)\cos(\mathbf{Q}\cdot\mathbf{r}+\phi)]\delta(\omega\mp\Delta_S)\\ &+[1+W(h)\cos(\mathbf{Q}\cdot\mathbf{r}+\phi)]\delta(\omega\mp\Delta_L).
\end{aligned}
\label{eq:weight}
\end{equation}
Here the weight function $W(h)\in[0,1]$ is a smooth monotonic function with $W(0)=0$. \Eq{eq:weight} can explain the emergence of 
coherence peak modulation below the ferromagnetic transition temperature \cite{zhao23n} in Eu-1144. If an external magnetic field is applied, the modulation should be visible above the FM transition temperature. Such a possibility can be verified in future experiments. 

On the other hand, the LDOS in the case where $s=-1$ (applicable to Bi-2212 and Bi-2223) read
\begin{equation}
\begin{aligned}
D_\pm^{s=-1}(\omega,\mathbf{r})&\sim[1\mp\cos(\mathbf{Q}\cdot\mathbf{r}+\phi)]\delta(\omega\mp\Delta_S)\\&+[1\pm\cos(\mathbf{Q}\cdot\mathbf{r}+\phi)]\delta(\omega\mp\Delta_L).
\end{aligned}
\end{equation}
Thus, the Zeeman field does not affect the particle-hole asymmetric coherence peak modulations, which can be checked experimentally by applying an in-plane field. 

In the experiments \cite{Zou2022,zhao23n}, the coherence peak modulations are observed to have finite coherence length. This is because the variations of the impurity scattering strength $V$ and the phase $\phi$ can broaden the coherence peaks, and lead to finite coherence length in the modulation. In fact, the coherence length of the modulation is contingent upon microscopic details of the impurity potential, in particular, the correlation length of the impurity potential itself. For instance, if the impurity potential exhibits a long correlation length, as is the case with out-of-plane impurities, the coherence length of modulation can be correspondingly long. Conversely, for impurities with short correlation lengths, such as point-like in-plane impurities, the coherence length of modulation will be shorter.

\section{Kagome metals $A$V$_3\textbf{Sb}_5$} The coherence peak modulations at three inequivalent wavevectors are observed in the kagome metal CsV$_3$Sb$_5$ \cite{chen21n}. The observed wavevectors coincide with the $3/4$ of the Bragg wavevectors. Interestingly, these are equivalent to the wavevectors connecting the three inequivalent small Fermi pockets after CDW orders reconstruct the Fermi surfaces \cite{lin21prb,lin22prb,zhou22nc,li23ax}. Importantly, the coherence peak modulation wavevectors are not observed in the normal phase above SC $T_c$. This suggests our PBSI scenario as a promising candidate for the origin of coherence peak modulations. Identifying the centers of the small Fermi pockets as the hot spots, the PBSI mechanism is again applicable. Note that the particle-hole symmetry (asymmetry) nature of the modulation is determined solely by the symmetry of the gap function and the impurity type. Since the nodeless $s$-wave pairing is proposed experimentally \cite{xu21prl}, we think the same-sign gap function at the hot spots is relevant. Meanwhile, the $\mu$SR experiments suggest the presence of local internal magnetic field \cite{yu21ax,guguchia23nc}, which can serve as the intrinsic Zeeman field and the source of magnetic scattering.

Based on these properties, we derive the particle-hole symmetric coherence peak modulations at three inequivalent wavevectors in S. M. Section III \cite{supp}. This result is consistent with the experiments \cite{chen21n,li23ax}.

It should be noted that for CsV$_3$Sb$_5$, \textit{one} CDW wavevector, which is 1/4 of a reciprocal lattice vector, occurs above SC $T_c$ \cite{zhao21nkm}. Since the 1/4 and the 3/4 wavevectors differed by a reciprocal lattice vector are equivalent, it is possible that the coherence peak modulation is induced by the CDW and and a uniform SC. However, we deem this as unlikely because there are \textit{three inequivalent} 3/4 wavevectors appearing in the coherence peak modulation. A similar STM experiment has been done on KV$_3$Sb$_5$ \cite{li23ax}, where, interestingly, the 1/4 wavevector is absent in the normal phase. In recent experiments, coherence peak modulations are observed in KV$_3$Sb$_5$ and CsV$_3$Sb$_5$ \cite{Deng2024Na,Deng2024NM}. Importantly, the 3/4-wavevector coherence peak modulations appear only when the impurities are dense enough, and they vanish in the defect-few regions. These results are in perfect agreement with the prediction of our PBSI mechanism.

\section{Conclusion} We have proposed the pair-breaking scattering interference (PBSI) as an alternative explanation of coherence peak modulations in superconductors. This explanation involves (i) the momentum locations of the hot spots, which can be obtained experimentally from, say, angle-resolved photoemission spectroscopy (ARPES); (ii) the sign structure of the gap function at the hot spots, which requires phase-sensitive measurements such as phase-sensitive quasiparticle interference STM spectroscopy; and (iii) the nature of impurities, which can be obtained through, say, NMR measurement and the phase-sensitive quasiparticle interference in STM. In particular, according to the PBSI mechanism, impurity scattering is essential. Hence one would expect coherence peak modulations to be detected around impurities. On the contrary, the signal should be absent in large impurity-free regions. The predictions of the pair-breaking scattering interference can be examined experimentally in STM measurements. This work serves as a cautious reminder to the scientific community when asserting the existence of a PDW phenomenon.

\begin{acknowledgments}
D.-H.L. thanks Yayu Wang for collaboration on Ref.~\cite{Zou2022} which motivates the current manuscript. He also thanks J. C. S\'eamus Davis for comments and the question concerning the modulation in the coherence peak height. D.-H.L. is grateful to Steve Kivelson for bringing the Eu-1144 experiment to his attention. Z.-Q.G. acknowledges Steve Kivelson, Srinivas Raghu, and Zhaoyu Han for fruitful discussions. This work is primarily funded by the U.S. Department of Energy, Office of Science, Office of Basic Energy Sciences, Materials Sciences and Engineering Division under Contract No. DE-AC02-05-CH11231 (Theory of Materials program KC2301). Y.-P.L. acknowledges the fellowship support from the Gordon and Betty Moore Foundation through the Emergent Phenomena in Quantum Systems (EPiQS) program.
\end{acknowledgments}

\bibliography{ref}

\begin{thebibliography}{23}
\bibitem [{not()}]{note1}%
  \BibitemOpen
  \bibinfo {note} {{This is indeed the case for Eu-1144 \cite{Zou2022}, since
  the Zeeman field originates from the weaker ferromagnetism in the stronger
  superconductivity. However, the results remain the same even if the reverse
  is true.}}\BibitemShut {Stop}%
\end{thebibliography}

\onecolumngrid

\renewcommand\theequation{S\arabic{equation}}
\renewcommand\thefigure{S\arabic{figure}}
\renewcommand\thetable{S\arabic{table}}
\renewcommand\bibnumfmt[1]{[S#1]}
\setcounter{equation}{0}
\setcounter{figure}{0}
\setcounter{table}{0}

\vspace{5cm}
\begin{center}
\Large{\bf Supplemental Material for ``Pair-breaking scattering interference as a mechanism for superconducting gap modulation"}
\end{center}

\section{The pair-breaking Hamiltonian}

In the solids, the electrons $c_{\mathbf r}=(c_{\mathbf r\uparrow},c_{\mathbf r\downarrow})^T$ can scatter by the scalar and magnetic impurities (for simplicity we assume the magnetic scattering to be $S_z$-conserving without losing generality)
\begin{equation}
H_\text{imp}=\sum_{\mathbf r}c_{\mathbf r}^\dag[V^\text{s}(\mathbf r)I+V^\text{m}(\mathbf r)Z]c_{\mathbf r},
\end{equation}
where $V^\text{s,m}(\mathbf r)$ are local scalar and magnetic impurity-scattering potentials, respectively. A Fourier transform $c_{\mathbf{r}}=(1/\sqrt{N})\sum_{\mathbf{k}}c_\mathbf{k}e^{i\mathbf{k}\cdot\mathbf{r}}$, where $N$ is the number of sites, yields the momentum-space form
\begin{equation}
H_\text{imp}=\sum_{\mathbf k\mathbf q}c_{\mathbf k}^\dag[V^\text{s}_\mathbf{q}I+V^\text{m}_\mathbf{q}Z]c_{\mathbf k-\mathbf{q}}.
\end{equation}
Here $V^\text{s,m}_\mathbf{q}=(1/N)\sum_\mathbf{r}V^\text{s,m}(\mathbf{r})e^{-i\mathbf{q}\cdot\mathbf{r}}$ are the Fourier-transformed impurity potentials, which are smooth in transfer momentum $\mathbf q$ due to their local structures. The expression in terms of the Nambu operator $\psi_{\bf k}=(c_{\bf k\uparrow},c_{\bf k\downarrow}, c^\dagger_{\bf -k\uparrow}, c^\dagger_{\bf -k\downarrow})^T$ reads
\begin{equation}
H_\text{imp}=\sum_{\mathbf k\mathbf q}\Psi_{\mathbf k}^\dag[V^\text{s}_\mathbf{q}ZI+V^\text{m}_\mathbf{q}ZZ]\Psi_{\mathbf k-\mathbf{q}}.
\end{equation}
Since $H_\text{imp}$ commutes with $ZZ$, we can decompose it into the $\uparrow$ ($ZZ=+1$) and $\downarrow$ ($ZZ=-1$) sectors with 
\begin{equation}
H_{\text{imp},\uparrow\downarrow}=\sum_{\mathbf k\mathbf q}\Psi_{\mathbf k\uparrow\downarrow}^\dag[V^\text{s}_\mathbf{q}Z\pm V^\text{m}_\mathbf{q}I]\Psi_{\mathbf k-\mathbf{q}\uparrow\downarrow},
\end{equation}
where $\psi_{\bf k\uparrow}=(c_{\bf k\uparrow}, c^\dagger_{\bf -k\uparrow})^T$ and $\psi_{\bf k\downarrow}=(c_{\bf k\downarrow},  c^\dagger_{\bf -k\downarrow})^T$.

Now we derive the pair-breaking Hamiltonian. Here we focus on the hot spots $\mathbf k_1$ and $\mathbf k_2$ on the Fermi surface, where the quasiparticle operators $\chi_{\mathbf{k}_{1,2}\uparrow\downarrow}$ and $\eta_{\mathbf{k}_{1,2}\uparrow\downarrow}$ are given by
\begin{equation}
\Psi_{\mathbf{k}_{1,2}\uparrow\downarrow}=
\frac{1}{\sqrt2}\left(\begin{array}{c}
\chi_{\mathbf{k}_{1,2}\uparrow\downarrow}+\eta_{\mathbf{k}_{1,2}\uparrow\downarrow}\\\pm e^{-i\theta_{\mathbf{k}_{1,2}}}[\chi_{\mathbf{k}_{1,2}\uparrow\downarrow}-\eta_{\mathbf{k}_{1,2}\uparrow\downarrow}]
\end{array}\right). 
\end{equation}
The impurity-scattering potential can be transformed into the quasiparticle eigenbasis
\begin{equation}
H_{\text{imp},\uparrow\downarrow}
=\frac{1}{\sqrt2}\left(\begin{array}{cc}
[\chi_{\mathbf{k}_1\uparrow\downarrow}^\dag+\eta_{\mathbf{k}_1\uparrow\downarrow}^\dag],&\pm e^{i\theta_{\mathbf{k}_1}}[\chi_{\mathbf{k}_1\uparrow\downarrow}^\dag-\eta_{\mathbf{k}_1\uparrow\downarrow}^\dag]
\end{array}\right)[V^\text{s}_\mathbf{Q}Z\pm V^\text{m}_\mathbf{Q}I]\frac{1}{\sqrt2}\left(\begin{array}{c}
\chi_{\mathbf{k}_2\uparrow\downarrow}+\eta_{\mathbf{k}_2\uparrow\downarrow}\\\pm e^{-i\theta_{\mathbf{k}_2}}[\chi_{\mathbf{k}_2\uparrow\downarrow}-\eta_{\mathbf{k}_2\uparrow\downarrow}]
\end{array}\right)+\text{h.c.}.
\end{equation}
The pair-breaking (PB) part of the Hamiltonian is obtained by omitting the crossing terms between $\chi_{\mathbf{k}_F\uparrow\downarrow}$ and $\eta_{\mathbf{k}_F\uparrow\downarrow}$, yielding
\begin{equation}
\begin{aligned}
H_{\text{PB},\uparrow\downarrow}
&=\chi_{\mathbf{k}_1\uparrow\downarrow}^\dag\left(\begin{array}{cc}
1,&\pm e^{i\theta_{\mathbf{k}_1}}
\end{array}\right)\frac{1}{2}\left(\begin{array}{cc}
V^\text{s}_\mathbf{Q}\pm V^\text{m}_\mathbf{Q}&0\\0&-V^\text{s}_\mathbf{Q}\pm V^\text{m}_\mathbf{Q}\end{array}\right)\left(\begin{array}{c}
1\\\pm e^{-i\theta_{\mathbf{k}_2}}
\end{array}\right)\chi_{\mathbf{k}_2\uparrow\downarrow}\\
&\quad+\eta_{\mathbf{k}_1\uparrow\downarrow}^\dag\left(\begin{array}{cc}
1,&\mp e^{i\theta_{\mathbf{k}_1}}
\end{array}\right)\frac{1}{2}\left(\begin{array}{cc}
V^\text{s}_\mathbf{Q}\pm V^\text{m}_\mathbf{Q}&0\\0&-V^\text{s}_\mathbf{Q}\pm V^\text{m}_\mathbf{Q}\end{array}\right)\left(\begin{array}{c}
1\\\mp e^{-i\theta_{\mathbf{k}_2}}
\end{array}\right)\eta_{\mathbf{k}_2\uparrow\downarrow}+\text{h.c.}\\
&=\chi_{\mathbf{k}_1\uparrow\downarrow}^\dag\frac{1}{2}[(V^\text{s}_\mathbf{Q}\pm V^\text{m}_\mathbf{Q})+(-V^\text{s}_\mathbf{Q}\pm V^\text{m}_\mathbf{Q})e^{i(\theta_{\mathbf{k}_1}-\theta_{\mathbf{k}_2})}]
\chi_{\mathbf{k}_2\uparrow\downarrow}\\
&\quad+\eta_{\mathbf{k}_1\uparrow\downarrow}^\dag\frac{1}{2}[(V^\text{s}_\mathbf{Q}\pm V^\text{m}_\mathbf{Q})+(-V^\text{s}_\mathbf{Q}\pm V^\text{m}_\mathbf{Q})e^{i(\theta_{\mathbf{k}_1}-\theta_{\mathbf{k}_2})}]\eta_{\mathbf{k}_2\uparrow\downarrow}+\text{h.c.}\\
&=\bigg\{V^\text{s}_\mathbf{Q}\bigg[\frac{1-e^{i(\theta_{\mathbf{k}_1}-\theta_{\mathbf{k}_2})}}{2}\bigg]\pm V^\text{m}_\mathbf{Q}\bigg[\frac{1+e^{i(\theta_{\mathbf{k}_1}-\theta_{\mathbf{k}_2})}}{2}\bigg]\bigg\}(\chi_{\mathbf{k}_1\uparrow\downarrow}^\dag
\chi_{\mathbf{k}_2\uparrow\downarrow}+\eta_{\mathbf{k}_1\uparrow\downarrow}^\dag\eta_{\mathbf{k}_2\uparrow\downarrow})+\text{h.c.}.
\end{aligned}\label{eq:S7}
\end{equation}
This is the pair-breaking Hamiltonian we use in the main text.

\section{Derivation of LDOS at zero temperature}

In the STM measurements, the LDOS at spatial position $\mathbf{r}$ and bias $\omega$ is determined by the local differential conductance
\begin{equation}
D(\omega,\mathbf{r})\sim\frac{dI}{dV}(\omega,\mathbf{r}),
\end{equation}
where $I$ and $V$ are the local current and voltage, respectively. At positive and negative biases, the differential conductance corresponding to the electron and the hole tunnelings are  
\begin{equation}
\begin{aligned}
D_+(\omega,\mathbf{r})&\sim\sum_n\sum_{\sigma}|\braket{n}{c_{\mathbf{r}\sigma}^\dag}{\Omega}|^2\delta(\omega-E_n),\\
D_-(\omega,\mathbf{r})&\sim\sum_n\sum_{\sigma}|\braket{n}{c_{\mathbf{r}\sigma}}{\Omega}|^2\delta(\omega+E_n).
\end{aligned}
\end{equation}
Here $\ket{\Omega}$ is the SC ground state and $\ket{n}$ are eigenstates with energies $E_n$. After Fourier transform, the tunneling amplitudes read
\begin{equation}
\begin{aligned}
\braket{n}{c_{\mathbf{r}\sigma}^\dag}{\Omega}&=\frac{1}{\sqrt{N}}\sum_\mathbf{k}\braket{n}{c_{\mathbf{k}\sigma}^\dag}{\Omega}e^{-i\mathbf{k}\cdot\mathbf{r}},\\
\braket{n}{c_{\mathbf{r}\sigma}}{\Omega}&=\frac{1}{\sqrt{N}}\sum_\mathbf{k}\braket{n}{c_{\mathbf{k}\sigma}}{\Omega}e^{i\mathbf{k}\cdot\mathbf{r}}.
\end{aligned}
\end{equation}
On the Fermi surface, we express the $c_{\mathbf{k}_F\sigma}$ in terms of the quasiparticle operators as
\begin{equation}
c_{\mathbf{k}_F\sigma}=\frac{1}{\sqrt{2}}(\chi_{\mathbf{k}_F\sigma}+\eta_{\mathbf{k}_F\sigma}).
\end{equation}
Note that the ground state obeys the condition $\chi_{\mathbf{k}_F\sigma}\ket{\Omega}=\eta_{\mathbf{k}_F\sigma}^\dag\ket{\Omega}=0$ (at zero temperature, the quasiparticle states above and below the Fermi level are fully filled and empty, respectively). The tunneling amplitudes then become
\begin{equation}
\begin{aligned}
\braket{n}{c_{\mathbf{k}_F\sigma}^\dag}{\Omega}&=\frac{1}{\sqrt{2}}\braket{n}{\chi_{\mathbf{k}_F\sigma}^\dag}{\Omega},\\
\braket{n}{c_{\mathbf{k}_F\sigma}}{\Omega}&=\frac{1}{\sqrt{2}}\braket{n}{\eta_{\mathbf{k}_F\sigma}}{\Omega}.
\end{aligned}
\end{equation}
Within the QP approximation, the relevant eigenstates $\ket{n}$ are the one-quasiparticle and one-quasihole excitations
\begin{equation}
\ket{n}=\sum_{\mathbf{k}_F}\sum_\sigma(\alpha_{n\mathbf{k}_F\sigma}^*\chi_{\mathbf{k}_F\sigma}^\dag+\beta_{n\mathbf{k}_F\sigma}\eta_{\mathbf{k}_F\sigma})\ket{\Omega}.
\end{equation}
According to the fact $\braket{\Omega}{\chi_{\mathbf{k}\sigma}\chi_{\mathbf{k}'\sigma'}^\dag}{\Omega}=\braket{\Omega}{\eta_{\mathbf{k}\sigma}^\dag\eta_{\mathbf{k}'\sigma'}}{\Omega}=\delta_{\mathbf{k}\mathbf{k}'}\delta_{\sigma\sigma'}$, the scattering amplitudes take the form
\begin{equation}
\begin{aligned}
\braket{n}{\chi_{\mathbf{k}_F\sigma}^\dag}{\Omega}&=\alpha_{n\mathbf{k}_F\sigma},\\
\braket{n}{\eta_{\mathbf{k}_F\sigma}}{\Omega}&=\beta_{n\mathbf{k}_F\sigma}^*.
\end{aligned}
\end{equation}
Inserting this result back to the real-space tunneling amplitudes, we get
\begin{equation}
\begin{aligned}
\braket{n}{c_{\mathbf{r}\sigma}^\dag}{\Omega}&=\frac{1}{\sqrt{2N}}\sum_{\mathbf{k}_F}\alpha_{n\mathbf{k}_F\sigma}e^{-i\mathbf{k}_F\cdot\mathbf{r}},\\
\braket{n}{c_{\mathbf{r}\sigma}}{\Omega}&=\frac{1}{\sqrt{2N}}\sum_{\mathbf{k}_F}\beta_{n\mathbf{k}_F\sigma}^*e^{i\mathbf{k}_F\cdot\mathbf{r}}.
\end{aligned}
\end{equation}

Now we derive the LDOS associated with the hot spots $\mathbf k_1$ and $\mathbf k_2$. The relevant tunneling amplitudes read
\begin{equation}
\begin{aligned}
\braket{n}{c_{\mathbf{r}\sigma}^\dag}{\Omega}&=\frac{1}{\sqrt{2N}}\left[\alpha_{n\mathbf{k}_1\sigma}e^{-i\mathbf{Q}\cdot\mathbf{r}}+\alpha_{n\mathbf{k}_2\sigma}\right]e^{-i\mathbf{k}_2\cdot\mathbf{r}},\\
\braket{n}{c_{\mathbf{r}\sigma}}{\Omega}&=\frac{1}{\sqrt{2N}}\left[\beta_{n\mathbf{k}_1\sigma}^*e^{i\mathbf{Q}\cdot\mathbf{r}}+\beta_{n\mathbf{k}_2\sigma}^*\right]e^{i\mathbf{k}_2\cdot\mathbf{r}},
\end{aligned}
\end{equation}
where ${\bf Q=k_1-k_2}$.
The LDOS then take the form
\begin{equation}
\begin{aligned}
D_+(\omega,\mathbf{r})&\sim\frac{1}{2N}\sum_n\sum_{\sigma}\left|\alpha_{n\mathbf{k}_1\sigma}e^{-i\mathbf{Q}\cdot\mathbf{r}}+\alpha_{n\mathbf{k}_2\sigma}\right|^2\delta(\omega-E_n),\\
D_-(\omega,\mathbf{r})&\sim\frac{1}{2N}\sum_n\sum_{\sigma}\left|\beta_{n\mathbf{k}_1\sigma}^*e^{i\mathbf{Q}\cdot\mathbf{r}}+\beta_{n\mathbf{k}_2\sigma}^*\right|^2\delta(\omega+E_n).
\end{aligned}
\end{equation}
The contributions from the four low-energy quasiparticle eigenstates $\chi_{\uparrow,S,L}$ and $\eta_{\downarrow,S,L}$ are
\begin{equation}
\begin{aligned}
D_+(\omega,\mathbf{r})&\sim\left|e^{-i\mathbf{Q}\cdot\mathbf{r}}-e^{i\phi}\right|^2\delta(\omega-\Delta_S)+\left|e^{-i\mathbf{Q}\cdot\mathbf{r}}+e^{i\phi}\right|^2\delta(\omega-\Delta_L)\\
&\sim[1-\cos(\mathbf{Q}\cdot\mathbf{r}+\phi)]\delta(\omega-\Delta_S)+[1+\cos(\mathbf{Q}\cdot\mathbf{r}+\phi)]\delta(\omega-\Delta_L),\\
D_-(\omega,\mathbf{r})&\sim\left|e^{i\mathbf{Q}\cdot\mathbf{r}}-se^{-i\phi}\right|^2\delta(\omega+\Delta_S)+\left|e^{i\mathbf{Q}\cdot\mathbf{r}}+se^{-i\phi}\right|^2\delta(\omega+\Delta_L)\\
&\sim[1-s\cos(\mathbf{Q}\cdot\mathbf{r}+\phi)]\delta(\omega+\Delta_S)+[1+s\cos(\mathbf{Q}\cdot\mathbf{r}+\phi)]\delta(\omega+\Delta_L).
\end{aligned}
\end{equation}
Similarly, we can do the calculation for the four high-energy quasiparticle eigenstates $\chi_{\downarrow,S,L}$ and $\eta_{\uparrow,S,L}$
\begin{equation}
\begin{aligned}
D_+'(\omega,\mathbf{r})&\sim\left|e^{-i\mathbf{Q}\cdot\mathbf{r}}+se^{i\phi}\right|^2\delta(\omega-\Delta_S')+\left|e^{-i\mathbf{Q}\cdot\mathbf{r}}-se^{i\phi}\right|^2\delta(\omega-\Delta_L')\\
&\sim[1+s\cos(\mathbf{Q}\cdot\mathbf{r}+\phi)]\delta(\omega-\Delta_S')+[1-s\cos(\mathbf{Q}\cdot\mathbf{r}+\phi)]\delta(\omega-\Delta_L'),\\
D_-'(\omega,\mathbf{r})&\sim\left|e^{i\mathbf{Q}\cdot\mathbf{r}}+e^{-i\phi}\right|^2\delta(\omega+\Delta_S')+\left|e^{i\mathbf{Q}\cdot\mathbf{r}}-e^{-i\phi}\right|^2\delta(\omega+\Delta_L')\\
&\sim[1+\cos(\mathbf{Q}\cdot\mathbf{r}+\phi)]\delta(\omega+\Delta_S')+[1-\cos(\mathbf{Q}\cdot\mathbf{r}+\phi)]\delta(\omega+\Delta_L').
\end{aligned}
\end{equation}
These are the LDOS formula we use in the main text.

\section{Superconducting gap modulations in $A\text{V}_3\text{Sb}_5$}

\begin{figure}[t]
  \centering
		\includegraphics[width=0.4\linewidth]{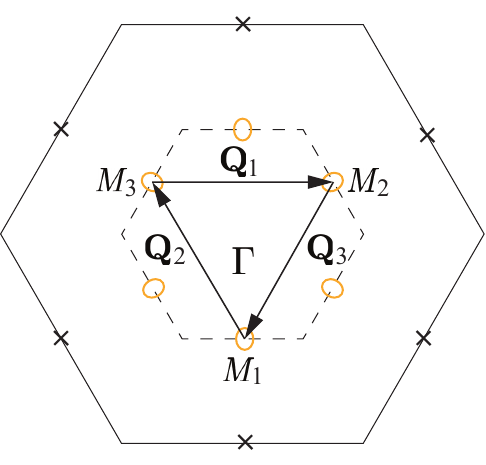}
		\caption{Small Fermi pockets (yellow) of kagome metals $A$V$_3$Sb$_5$ at $M_{1,2,3}$ points. The gap function has the same sign on all these pockets. The pockets at opposite momenta are equivalent, since they are connected by a reciprocal lattice vector of the folded BZ. The three inequivalent Fermi pockets are connected by the wavevectors $\mathbf{Q}_{1,2,3}$. The Bragg peaks of the CDW phase are marked by black crosses. For clarity, the wavevectors of the CDW order connecting these peaks are omitted. }\label{fig:sm1}
\end{figure}

The formalism developed in this work can be also extended to systems with more than two hot spots, which as shown below yields commendable agreement with the experiments. In the kagome metals $A$V$_3$Sb$_5$, the $2\times2$ CDW order folds the BZ into the $1/2\times1/2$ one (the area enclosed by the dashed lines in Fig.~\ref{fig:sm1}). The reciprocal lattice vectors of the folded BZ connect the origin and the black crosses in Fig.~\ref{fig:sm1}, with the CDW ordering wavevectors connecting the black crosses. After the CDW-induced band reconstruction, small Fermi pockets appear at three inequivalent momenta $M_{1,2,3}$ on the edges of the folded BZ. The inequivalent Fermi pockets are connected by three wavevectors $\mathbf{Q}_{1,2,3}$. These wavevectors are $1/4$ of the original reciprocal lattice vectors. They are related to the modulation wavevectors of the coherence peak, modulo the receiprocal lattice vectors. It is important to note that $\mathbf{Q}_{1,2,3}$ are distinct from the CDW ordering wavevectors. In the following we identify the centers of the small Fermi pockets as the hot spots. They give rise to the coherence peak modulations wavevectors in the superconducting state via the PBSI mechanism.  

Our analysis assumes the same-sign gap functions at all hot spots. According to the main text, the relevant pair-breaking impurity is the magnetic one. Similar to Eq.(5) in the main text and \Eq{eq:S7} in the previous section, the pair-breaking part of the impurity scattering Hamiltonian is
\begin{equation}
H_{\text{PB},\uparrow,\downarrow}
=\chi_{\uparrow\downarrow}^\dagger \mathcal H_{\text{PB},\uparrow,\downarrow}\chi_{\uparrow\downarrow}+\eta_{\uparrow\downarrow}^\dagger \mathcal H_{\text{PB},\uparrow,\downarrow}\eta_{\uparrow\downarrow}.
\end{equation}
Here $\chi_{\uparrow\downarrow}=(\chi_{M_1\uparrow\downarrow},\chi_{M_2\uparrow\downarrow},\chi_{M_3\uparrow\downarrow})^T$ and $\eta_{\uparrow\downarrow}=(\eta_{M_1\uparrow\downarrow},\eta_{M_2\uparrow\downarrow},\eta_{M_3\uparrow\downarrow})^T$ represent the 3-component quasiparticle operators at the hot spots $M_{1,2,3}$, and
\begin{equation}
\mathcal H_{\text{PB},\uparrow,\downarrow}=\pm \begin{pmatrix}
    0 & V^\text{m}_{\mathbf{Q}_3} & (V^\text{m}_{\mathbf{Q}_2})^\dagger\\
    (V^\text{m}_{\mathbf{Q}_3})^\dagger & 0 &  V^\text{m}_{\mathbf{Q}_1}\\
    V^\text{m}_{\mathbf{Q}_2} & (V^\text{m}_{\mathbf{Q}_1})^\dagger & 0
\end{pmatrix}.
\end{equation}
Due to the involvement of three hot spots, the energy splitting is more complicated than the 2-hot-spot results. Here we consider a simplified case where the $C_3$ rotation symmetry is preserved. The pair-breaking Hamiltonian can be chosen, for example, as
\begin{equation}
\mathcal H_{\text{PB},\uparrow,\downarrow}=\pm V\begin{pmatrix}
    0 & e^{i\phi} & e^{-i\phi}\\
    e^{-i\phi} & 0 &  e^{i\phi}\\
    e^{i\phi} & e^{-i\phi} & 0
\end{pmatrix}.
\end{equation}
As in the 2-hot-spot case, we focus on the lowest-energy branches $\chi_\uparrow$ and $\eta_\downarrow$. The eigenstates and corresponding energies are
\begin{equation}
\begin{aligned}
E_{+n}(\phi)&=(\Delta_0-h)+2V\cos(\phi+2\pi n/3),\quad\chi_{\uparrow n}=\frac{1}{\sqrt3}(e^{i2\pi n/3}\chi_{\mathbf M_1\uparrow}+e^{-i2\pi n/3}\chi_{\mathbf M_2\uparrow}+\chi_{\mathbf M_3\uparrow}),\\
E_{-n}(\phi)&=-(\Delta_0-h)-2V\cos(\phi+2\pi n/3),\quad\eta_{\downarrow n}=\frac{1}{\sqrt3}(e^{i2\pi n/3}\eta_{\mathbf M_1\downarrow}+e^{-i2\pi n/3}\eta_{\mathbf M_2\downarrow}+\eta_{\mathbf M_3\downarrow}),
\end{aligned}
\end{equation}
where $n=0,1,2$ labels the eigenstates, $\Delta_0$ is the average superconducting gap, and $h$ is the Zeeman field. Note that the three-fold degeneracy is lifted by the pair-breaking scattering.

\begin{figure}[t]
  \centering
		\includegraphics[width=0.8\linewidth]{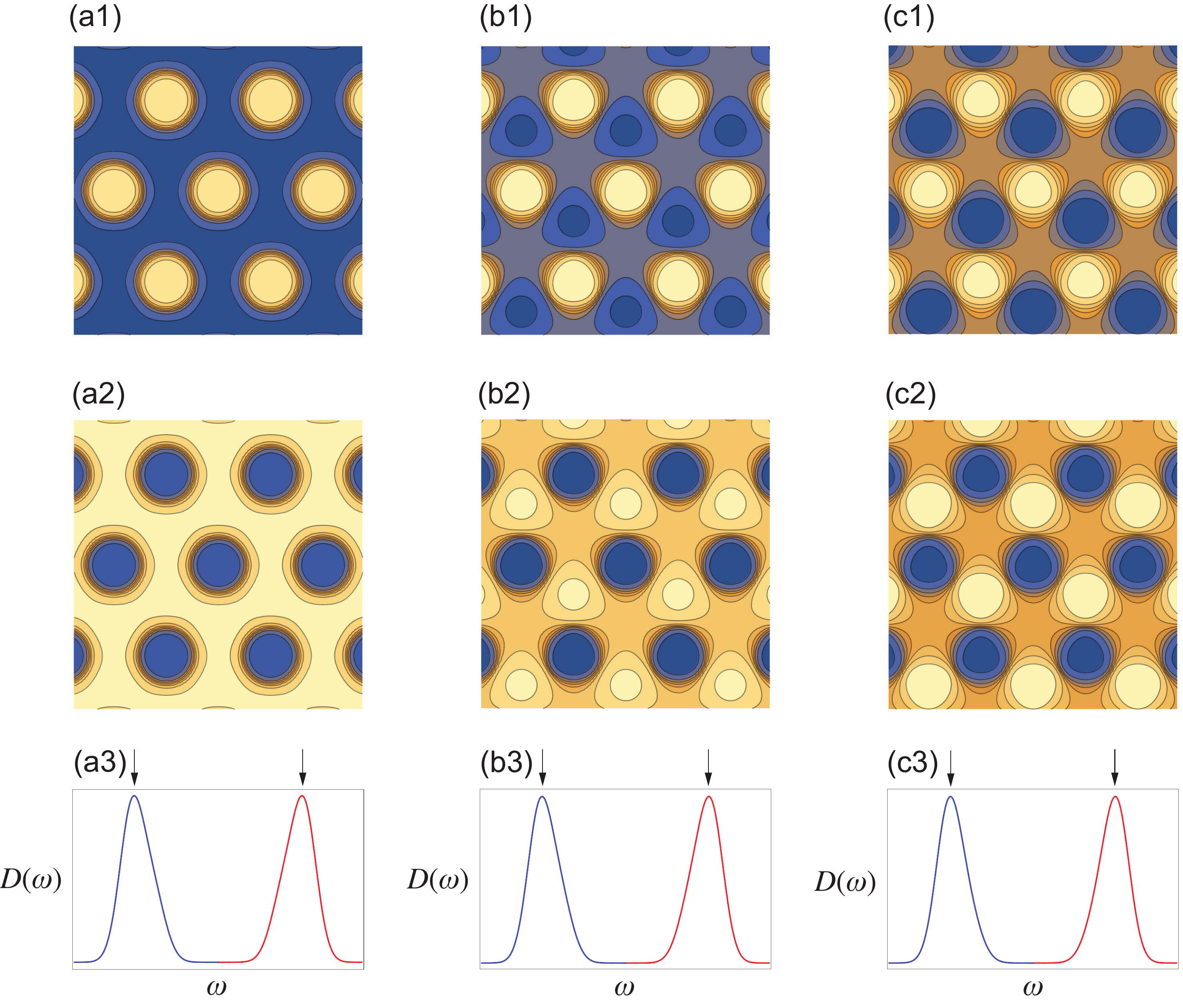}
		\caption{The energy separation of the coherence peaks as a function of spatial coordinates at the positive bias $\Delta_+(\mathbf{r},\phi)$ with (a1) $\phi=0.0$, (b1) $\phi=0.2$, and (c1) $\phi=0.4$, and at the negative bias $\Delta_-(\mathbf{r},\phi)$ with (a2) $\phi=0.0$, (b2) $\phi=0.2$, and (c2) $\phi=0.4$. These plots are calculated in the presence of a Zeeman field, since the kagome materials fall into the $s=1$ category in the main text and hence coherence-peak modulations only occur when Zeeman field is non-zero. It can be observed that the change of $\phi$ does not affect the wavevectors of the modulations which is particle-hole symmetric. Illustration of the superconducting gaps as the first local maxima of $D_+(\omega,\mathbf{r})$ (red) and $D_-(\omega,\mathbf{r})$ (blue) for a fixed $\mathbf{r}$ with (a3) $\phi=0.0$, (b3) $\phi=0.2$ and (c3) $\phi=0.4$, respectively, where the first maxima are pointed by arrows.}\label{fig:sm2}
\end{figure}

The LDOS can be derived similarly to the 2-hot-spot case. From the tunneling amplitudes
\begin{equation}
\begin{aligned}
|\braket{n}{c_{\mathbf{r}\sigma}^\dag}{\Omega}|^2&=\frac{1}{2N}|\alpha_{n\mathbf{M}_1\sigma}e^{-i\mathbf{M}_1\cdot\mathbf{r}}+\alpha_{n\mathbf{M}_2\sigma}e^{-i\mathbf{M}_2\cdot\mathbf{r}}+\alpha_{n\mathbf{M}_3\sigma}e^{-i\mathbf{M}_3\cdot\mathbf{r}}|^2\\
&=\frac{1}{2N}\frac{1}{3}|e^{i2\pi n/3}e^{-i\mathbf{M}_1\cdot\mathbf{r}}+e^{-i2\pi n/3}e^{-i\mathbf{M}_2\cdot\mathbf{r}}+e^{-i\mathbf{M}_3\cdot\mathbf{r}}|^2\\
&\sim3+2\sum_{\alpha=1}^3\cos(\mathbf{Q}_\alpha\cdot\mathbf{r}+2\pi n/3),\\
|\braket{n}{c_{\mathbf{r}\sigma}}{\Omega}|^2&=\frac{1}{2N}|\beta_{n\mathbf{M}_1\sigma}^*e^{i\mathbf{M}_1\cdot\mathbf{r}}+\beta_{n\mathbf{M}_2\sigma}^*e^{i\mathbf{M}_2\cdot\mathbf{r}}+\beta_{n\mathbf{M}_3\sigma}^*e^{i\mathbf{M}_3\cdot\mathbf{r}}|^2\\
&=\frac{1}{2N}\frac{1}{3}|e^{-i2\pi n/3}e^{i\mathbf{M}_1\cdot\mathbf{r}}+e^{i2\pi n/3}e^{i\mathbf{M}_2\cdot\mathbf{r}}+e^{i\mathbf{M}_3\cdot\mathbf{r}}|^2\\
&\sim3+2\sum_{\alpha=1}^3\cos(\mathbf{Q}_\alpha\cdot\mathbf{r}+2\pi n/3),
\end{aligned}
\end{equation}
we obtain the corresponding LDOS
\begin{equation}
\begin{aligned}
D_{\pm}(\omega,\mathbf{r},\phi)&\sim\sum_{n=0}^2\left[3+2\sum_{\alpha=1}^3\cos(\mathbf{Q}_\alpha\cdot\mathbf{r}+2\pi n/3)\right]\delta\left(\omega\mp[(\Delta_0-h)+ 2V\cos(\phi+2\pi n/3)]\right).
\end{aligned}
\end{equation}
Note that the result is particle-hole symmetric, as expected for the situation of same-sign gap functions under magnetic impurity scattering. The spatially varying superconducting gap, i.e. the energy position of coherence peaks, $\Delta_\pm(\mathbf{r},\phi)$ is determined by the first local maxima of LDOS (see Figs.~\ref{fig:sm2} for illustration)
\begin{equation}
\left.\frac{\partial}{\partial\omega}D_{\pm}(\omega,\mathbf{r},\phi)\right|_{\omega=\Delta_\pm(\mathbf{r},\phi)}=0.
\end{equation}
Without loss of generality, we restrict $\phi\in [0,\frac{\pi}{6})$. The profile of positive- and negative-bias superconducting gap functions $\Delta_\pm (\mathbf{r},\phi)$ is plotted for $\phi=0.0$, $0.2$, and $0.4$ in Figs.~\ref{fig:sm2}, where the modulations show the $C_3$ symmetry explicitly. Notably, the wavevectors $\mathbf{Q}_{1,2,3}$ are not affected by the phase $\phi$, i.e. the details of the impurity scattering. Although the modulation profile changes for different $\phi$, the particle-hole symmetry and the wavevectors remain unchanged.

\end{document}